\documentclass[aps,preprint,groupedaddress,showkeys]{revtex4}
\usepackage{amsmath}
\usepackage{graphicx}
\begin{document}


\title{{\it {RXJ1856.5-3754}} and {\it {RXJ0720.4-3125}} are P-Stars}


\author{Paolo Cea$^{1,2,}$}
\email[]{Paolo.Cea@ba.infn.it}
\affiliation{$^1$Physics Department, Univ. of Bari, I-70126 Bari, Italy \\
$^2$INFN - Sezione di Bari, I-70126 Bari, Italy}

%
\begin{abstract}
P-stars are a new class of compact stars made of up and down
quarks in $\beta$-equilibrium with electrons in a chromomagnetic
condensate. P-stars are able to account for  compact stars with $R
\,  \lesssim 6 \, Km$, as well as stars  comparable to canonical
neutron stars. We show that P-stars once formed are absolutely
stable, for they  cannot decay into neutron or strange stars. We
convincingly argue that the nearest isolated compact stars {\it
{RXJ1856.5-3754}} and {\it {RXJ0720.4-3125}} could be interpreted
as P-stars with $M \, \simeq 0.8 \, M_{\bigodot}$ and $R \, \simeq
5 \, Km$.

\end{abstract}

\keywords{Compact Star, Pulsar, Magnetic Field}

\maketitle

%
%
\renewcommand{\thesection}{\normalsize{{\arabic{section}}}}
\section{\normalsize{INTRODUCTION}}
\renewcommand{\thesection}{\arabic{section}}
Soon after the first radio pulsar were
discovered~\cite{hewish:1968}, it becomes generally accepted that
pulsars are rapidly rotating neutron stars endowed with a strong
magnetic field~\cite{pacini:1968,gold:1968}. The exact mechanism
by which a pulsar radiates the energy observed as radio pulses is
still a subject of vigorous debate~\cite{michel:1982,michel:1991},
nevertheless the accepted standard model based on the picture of a
rotating magnetic dipole has been developed since long
time~\cite{goldreich:1969,sturrock:1971}.
Nowadays, no one doubts that pulsars are indeed neutron stars.
However, quite recently we have proposed~\cite{cea:2003} a new
class of compact stars, named P-stars, which is challenging the
two pillars of modern astrophysics, namely neutron stars and black
holes. Indeed, in Ref.~\cite{cea:2003} we showed that P-stars,
which are compact stars made of up and down quarks in
$\beta$-equilibrium with electrons in an abelian chromomagnetic
condensate, are able to account for compact stars as well as stars
comparable to canonical neutron stars. Moreover, for stars with
mass $ M \, \simeq \, 1.4 \, M_{\bigodot}$ we
found~\cite{cea:2003} that the binding energy per nucleon is:
\begin{equation}
\label{bind}
B \; \simeq \; 0.386 \; GeV \; \; \;  , \; \; \;  A \simeq 2.9 \,
10^{57} \; \; \; ,
\end{equation}
where $A$ is the baryon number of the star. This result should be
compared with the binding energy per nucleon in the case of
limiting mass neutron star~\cite{glendenning:2000}, $B_{NS} \simeq
100 \; MeV$, showing that P-stars are gravitationally more bounded
than neutron stars. Moreover, it turns out that P-stars are also
more stable than strange stars~\cite{alcock:1986,haensel:1986}.
Indeed, it is known that there is extra energy of about $20 \;
MeV$ per baryon available from the conversion of matter to strange
matter~\cite{alcock:1986}. So that, even the binding energy per
nucleon for strange stars, $B_{SS} \simeq 120 \; MeV$, is much
smaller than the value in Eq.~(\ref{bind}). This shows that
gravitational effects make two flavour quark matter in the
chromomagnetic condensate  globally more stable than nuclear
matter or three flavour quark matter. However, one could wonder if
locally, let's say in a small volume of linear size $a \; \simeq
1/m_\pi \; \simeq \; 1 \; fermi$, two flavor quark matter immersed
in the chromomagnetic condensate could convert into neutrons or
deconfined quarks of relevance for strange stars. In order to see
how far this conversion will proceed, we need to estimate the
appropriate transition rates. Let us first consider the decay into
nuclear matter. In this case we must rearrange three quarks, which
are in the lowest Landau levels, into configurations of quarks in
nucleons. In doing this, we get a first suppression factor
$\alpha$ due to color mismatch. In fact, the quarks into the
lowest Landau levels are colorless due to presence of the
chromomagnetic condensate. This suppression factor cannot be
easily evaluated. However, in general we have that $\alpha \;
\lesssim \; 1$. Another suppression factor arises from the
mismatch of the quark wavefunctions in the directions transverse
to the chromomagnetic field. This results in a factor $\left
(\frac{m_{\pi}}{\sqrt{gH}} \right )^2$ for each quarks. Finally,
to arrange three quarks into a nucleon we need to flip one quark
spin, which increases the energy by a factor $\sqrt{gH}$ at least.
So that we have for the transition rate:
\begin{equation}
\label{prob}
P \; \; \sim \; \; \alpha \; \left (\frac{m_{\pi}}{\sqrt{gH}}
\right )^6 \; \; e^{-\frac{\sqrt{gH}}{T}} \; \; ,
\end{equation}
where $T$ is the star core temperature. For a newly born compact
star the core temperature is of order of several $MeV$ and it is
rapidly decreasing. On the other hand, the typical values of
$\sqrt{gH}$ is of order of several hundred of $MeV$. As a
consequence we have:
\begin{equation}
\label{prob_num}
P \; \; \lesssim \; \; 1.5 \; \; 10^{-46} \; \;,
\end{equation}
which implies that the decays of two flavor quark matter  immersed
in the chromomagnetic condensate into nuclear matter is
practically never realized. In addition, the conversion into
three-flavor quarks is even more suppressed. Indeed, first we need
to flip a large number of quark spin to fill the Fermi sphere,
then the quarks at the top of the Fermi sphere are allowed to
decay into strange quarks. We conclude, thus, that P-stars, once
formed, are absolutely stable.
The logical consequence is that now  we must admit that  supernova
explosions give rise to P-stars. In other words, we are lead to
identify pulsars with P-stars instead of neutron or strange stars.
Such a dramatic change in the standard paradigm of relativistic
astrophysics has been already advanced in our previous
paper~\cite{cea:2003} where we showed that, if we assume that
pulsars are P-stars, then we may completely solve the supernova
explosion problem. Indeed, the binding energy is the energy
released when the core of an evolved massive star collapses.
Actually, only about one percent of the energy appears as kinetic
energy in the supernova explosion~\cite{bethe:1990}. From
Eq.~(\ref{bind}) it follows that there is an extra gain in kinetic
energy of about $ \, 1 \, - \, 10 \, foe$ ($1 \, foe \, = \,
10^{51} \, erg$), which is enough to fire the supernova
explosions. Moreover, further support to our point of view comes
from cooling properties of P-stars. In fact, we found that cooling
curves of P-stars compare rather well with available observational
data. We are, however, aware that such a dramatic change in the
standard paradigm of relativistic astrophysics which is based on
neutron stars and black holes needs a careful comparison with the
huge amount of observations collected so far  for pulsar and black
hole candidates.
In a forthcoming paper~\cite{cea:2004} we discuss the generation
of magnetic field and the glitch mechanism in P-stars. In
particular, we find that for P-star with canonical mass $ M \,
\simeq \, 1.4 \, M_{\bigodot}$ our theory allows dipolar surface
magnetic fields up to about $B_S \, \simeq \, 10^{17} \; Gauss$.
Moreover, it turns out that the magnetic field is proportional to
the square of the spin period:
\begin{equation}
\label{magn-period}
B_S\; \; \simeq \; \; B_1 \; \left (\frac{P}{1 \, sec } \right )^2
\; \; \; \; ,
\end{equation}
where $B_1$ is the surface magnetic field for pulsars with period
$P \, = \, 1 \; sec$. Remarkably, assuming $B_1 \, \simeq \, 1.3
\, 10^{13} \; Gauss$, we find the Eq.~(\ref{magn-period}) accounts
rather well the inferred magnetic field for pulsars ranging from
millisecond pulsars up to anomalous $X$-ray pulsars and soft-gamma
repeaters. As a consequence of Eq.~(\ref{magn-period}), we see
that the dipolar magnetic field is time dependent. In fact, it is
easy to find:
\begin{equation}
\label{mag-time}
 B_S(t) \; \simeq \;  B_0 \; \; \left ( 1 \; + \; 2 \; \frac{\dot{P}}{P} \; t \;  \right )\; \;  \; \; ,
\end{equation}
where $B_0$ indicates the magnetic field at the initial
observation time. Note that Eq.~(\ref{mag-time}) implies that the
magnetic field varies on a time scale given by the characteristic
age:
\begin{equation}
\label{char-age}
 \tau_c \; = \; \frac{P}{2 \, \dot{P}} \; \;.
\end{equation}
A remarkable consequence of Eq.~(\ref{mag-time}) is that the
effective braking index $n$ is time dependent. In particular, the
braking index decreases with time such that:
\begin{equation}
\label{braking}
 -1 \; \lesssim \; n \; \lesssim \; 3  \; \;,
\end{equation}
the time scale variation being of order of $\tau_c/2$. However, it
turns out that~\cite{cea:2004} the monotonic derive of the braking
index is contrasted by the glitch activity. Indeed, in our theory
the glitches originate from dissipative effects in the inner core
of the star leading to a decrease of the strength of the dipolar
magnetic field, but to an increase of the magnetic torque.
Moreover, we find that the time variation of the dipolar magnetic
field is the origin of pulsar timing noise.

Concerning black holes, we feel that the most interesting and
intriguing aspect of our  theory is that P-stars do not admit the
existence of an upper limit to the mass of a completely degenerate
configuration. In other words, our peculiar equation of state of
degenerate up and down quarks in a chromomagnetic condensate
allows the existence of finite equilibrium states for stars of
arbitrary mass.  In a future publication we shall address the
experimental evidence for massive P-star. In particular, we shall
discuss the so-called galactic black hole candidates and $SgrA^*$,
the super massive compact object at the galactic center.

The aim of the present paper is to discuss in details the nearest
isolated radio quiet compact stars. In the next Section we briefly
review the theory of P-stars. After that, we discuss explicitly
the  isolated compact stars {\it {RXJ1856.5-3754}} and {\it
{RXJ0720.4-3125}}. From the emission spectrum we argue that the
most realistic interpretation is that these objects are compact
P-stars with $M \, \simeq 0.8 \, M_{\bigodot}$ and $R \, \simeq 5
\, Km$. Finally, our conclusions are drawn in Section~3.
\renewcommand{\thesection}{\normalsize{{\arabic{section}}}}
\section{\normalsize{ RXJ1856.5-3754 and RXJ0720.4-3125 }}
\begin{figure}
\includegraphics[width=0.90\textwidth,clip]{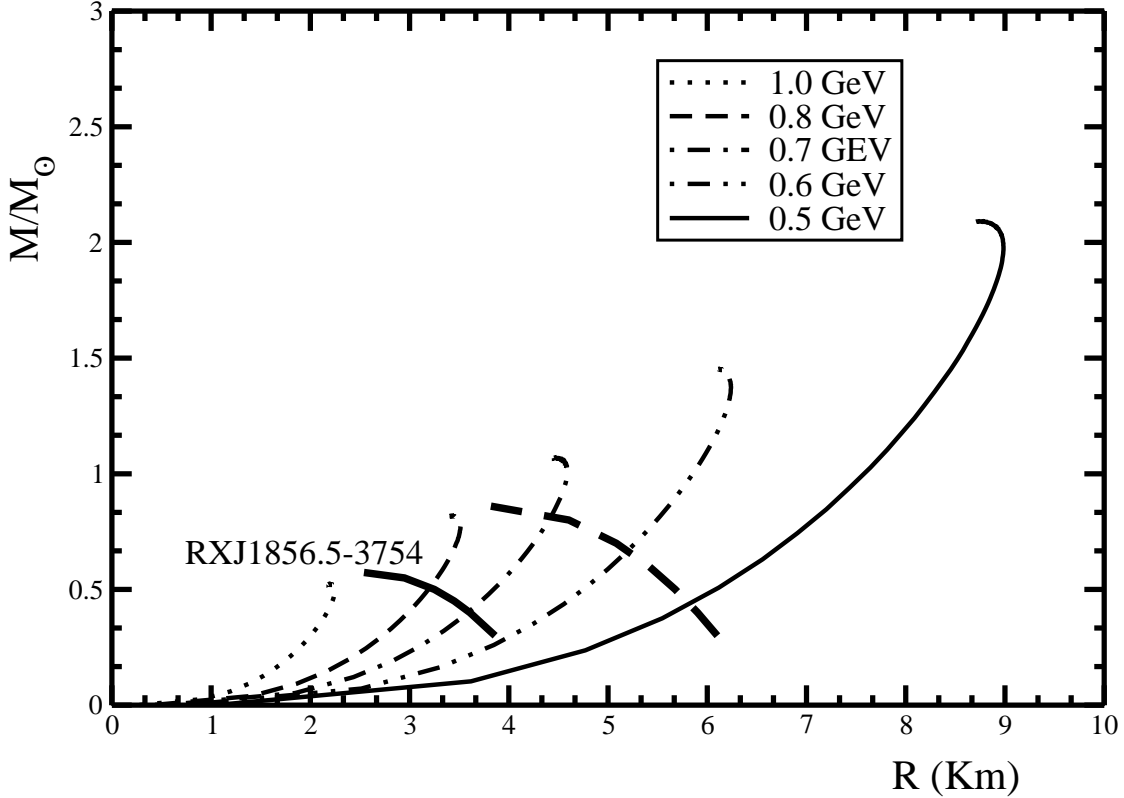}
\caption{\label{fig:01}
Gravitational mass M plotted versus stellar radius R for P-stars
for different values of $\sqrt{gH}$. Thick solid and dashed lines
correspond to M-R curve for {\it {RXJ1856}} obtained solving
Eq.~(\ref{radiuseff}) with $R^\infty = 4.4 \; Km$ and $R^\infty =
6.6 \; Km$ respectively, and assuming that $R \geq R_{\gamma} = 3
G M$ .}
\end{figure}

\renewcommand{\thesection}{\arabic{section}}
In our previous paper~\cite{cea:2003} we showed that P-stars are
able to account for compact stars with $R \, \lesssim 6 \, Km$, as
well as stars comparable to canonical neutron stars. In this
Section, after reviewing  P-stars, we will focus on the nearest
isolated compact stars {\it {RXJ1856.5-3754}} and {\it
{RXJ0720.4-3125}} (henceforth {\it {RXJ1856}} and {\it {RXJ0720}}
respectively).  We will convincingly  argue that these compact
stars are indeed P-stars.
P-stars are rather different from strange
stars~\cite{alcock:1986,haensel:1986} which are made entirely of
deconfined  u, d, s quarks. The possible existence of strange
stars is a direct consequence of the so called strange matter
hypothesis, namely that  u, d, s quarks in equilibrium with
respect to weak interactions could be the true ground state of
hadronic matter~\cite{bodmer:1971,witten:1984,farhi:1984}. Indeed,
as discussed in Section~1,  P-stars are more stable than neutron
stars and strange stars whatever the value of  the strength of the
chromagnetic condensate.
To investigate the structure of P-stars we  needs the equation of
state appropriate for the description of deconfined quarks and
gluons in an abelian chromomagnetic field. In general, the quark
chemical potentials turn out to be smaller that the strength of
the chromomagnetic field  $\sqrt{gH}$. As a consequence  up and
down quarks are in the lowest Landau levels with energy
$\varepsilon_{0,p_3} = |p_3|$, where $\vec{p}$ is the quark
momentum. It is straightforward  to determine the thermodynamic
potential (at $T \, = \, 0$). The energy density is given by:
\begin{equation}
\label{equark}
 \varepsilon \; = \; \frac{1}{4 \pi^2} \, gH \,
 ( \mu_u^2 + \mu_d^2 )  \, + \, \frac{\mu_e^4 }{4 \pi^2} \,
 + \, \frac{11}{32 \pi^2} \, (gH)^2
 \, ,
\end{equation}
where $\mu_f$ ( $f = d,u,e$) denotes the chemical potential. On
the other hand, we find for the pressure P:
\begin{equation}
\label{press}
 P \; = \; \frac{1}{4 \pi^2} \, gH \, ( \mu_u^2 +
\mu_d^2 ) \, + \, \frac{\mu_e^4 }{12 \pi^2}
 \, - \,  \frac{11}{32 \pi^2} \, (gH)^2 \, .
\end{equation}
Assuming that the system is in equilibrium with respect to weak
$\beta$-decays we have the following constraint:
\begin{equation}
\label{constr1}
 \mu_e \, + \, \mu_u \; = \; \mu_d  \; \; \; \; \;
\; \;  .
\end{equation}
Moreover from charge neutrality we also have:
\begin{equation}
\label{constr2}
 \frac{2}{3 } n_u \, - \, \frac{1}{3 } n_d  \; = \;
n_e  \; \; \; \; ,
\end{equation}
where
\begin{equation}
\label{number}
  n_u \, = \, \frac{1}{2 \pi^2} \, gH \, \mu_u  \,
,  \;  n_d \, = \, \frac{1}{2 \pi^2} \, gH \, \mu_d  \, , \;  n_e
\, = \, \frac{\mu_e^3}{3 \pi^2} \, .
\end{equation}
From previous equations one can easily write down the equation of
state:
\begin{equation}
\label{eos}
 P \; = \; \varepsilon \;  -  \;  \frac{\mu_e^4 }{6
\pi^2}
 \; - \;  \frac{11}{16 \pi^2} \, (gH)^2 \, .
\end{equation}
It is interesting to observe that the sound speed:
\begin{equation}
\label{sound}
 v^2_S \, = \, \frac{d P}{d \varepsilon } \, = \,1 \, -  \;
\frac{1}{\frac{39}{2} \, + \, 18 \, \overline{\mu}^2_e \, + \,
\frac{15}{4 \, \overline{\mu}^2_e}} \; ,
\end{equation}
is quite close to the causal limit $v_S \, = \, 1$, for
$\overline{\mu}_e \, = \, \mu_e / \sqrt{gH} \, < \, 1 $, leading
to a very stiff equation of state.
To study the properties of the star we consider it to be a
spherical symmetric object, corresponding to a non rotating star.
The stability of the star is governed by the general-relativistic
equation of hydrostatic equilibrium for a spherical configuration
of quark matter, which is the Tolman-Oppenheimer-Volkov
equation~\cite{glendenning:2000}:
\begin{equation}
\label{TOV}
 \frac{d P}{d r} \, = \, - \; \frac{G M(r)
\varepsilon(r)}{r^2} \,[ 1 \, + \, \frac{4 \pi r^3 P(r)}{M(r)}] \;
\frac{1 \, + \,\frac{P(r)}{\varepsilon(r)} }{1 \, - \, \frac{2
GM(r)}{r}} \; ,
\end{equation}
\begin{equation}
\label{TOV1}
 \frac{d M}{d r} \, = \, 4 \; \pi \; r^2 \:
\varepsilon(r) \; .
\end{equation}
These equations can be solved numerically for a given central
density $\varepsilon_c$. In this way we obtain $\varepsilon(r)$
and $P(r)$. The radius of the star is determined by :
\begin{equation}
\label{radius}
 P(r=R) \; \; = \; \; 0 \; \; ,
\end{equation}
while the total mass by:
\begin{equation}
\label{mass}
  M \; \; =  \; \; M(r=R) \; \; \; .
\end{equation}
In Figure~\ref{fig:01} we display the gravitational mass $M$
versus the star radius $R$ for different values of $\sqrt{gH}$. At
fixed value of the chromomagnetic condensate the mass and radius
of the star can be thought of as function of the central energy
density $\varepsilon_c$. We see that, as in  strange stars, the
mass first increases with $\varepsilon_c$ until it reaches a
maximum such that $\frac{d M(\varepsilon_c)}{d \varepsilon_c} \, =
\, 0$. The further increase of $\varepsilon_c$ leads in a region
where $\frac{d M(\varepsilon_c)}{d \varepsilon_c} \, < \, 0$ and
the system becomes unstable. Moreover, we see that there is no
lower limit for the radius R. Indeed, for small mass we find:
\begin{equation}
\label{radiuslow}
 R\, \sim \, M^{\frac{1}{3}} \; .
\end{equation}
Interestingly enough, Fig.~\ref{fig:01} shows that there are
stable P-stars with $M \, \lesssim \, M_{\bigodot}$ and $R \,
\lesssim \, 6 \, Km$. This region of the $M-R$ plane could be
relevant for the recently observed compact star {\it
{RXJ1856}}~\cite{walter:1996,neuhauser:1997}.
{\it {RXJ1856}} is the nearest and brightest of a class of
isolated radio-quiet compact stars (for a recent review
see~\cite{haberl:2003,pavlov:2003}). We shall also consider {\it
{RXJ0720}}, the second brightest isolated compact star.
{\it {RXJ1856}} has been observed with Chandra and
XMM-Newton~\cite{burwitz:2002}, showing that the $X$-ray spectrum
is accurately fitted by a blackbody law. Assuming that the $X$-ray
thermal emission  is due to the surface of the star, the authors
of Ref.~\cite{burwitz:2002} found for the effective radius and
surface temperature:
\begin{equation}
\label{rad-temp}
 R^\infty \, \simeq \, 4.4 \, \frac{d}{120 \, pc} \, Km \; \; , \; \; T^\infty \, \simeq
 \, 63 \, eV \; ,
\end{equation}
where
\begin{equation}
\label{radiuseff}
 R^\infty \, = \, \frac{R}{\sqrt{ 1 \, - \frac{2GM}{ R}}} \; , \;
 T^\infty \, = \, T \, \sqrt{ 1 \, - \frac{2GM}{ R}} \; .
\end{equation}
Assuming $ R^\infty = 4.4 \, Km$ (i.e. $ d = 120 \, pc$) we can
solve Eq.~(\ref{radiuseff}) for the true radius $R$. We also
impose that $ R \, \geq \, R_\gamma \, = \, 3 G M$, for it is well
known that the last circular photon orbit for the Schwarzschild
geometry is at $R_\gamma $. Remarkably, Fig.~1 shows that there
are stable P-star configurations which agree with observation
data. However, it should be stressed that in the observed spectrum
there is also an optical emission  in excess over the extrapolated
$X$-ray blackbody. By interpreting the optical emission as a
Rayleigh-Jeans tail of a thermal blackbody emission, one finds
that the optical data are also fitted by the blackbody model
yielding an effective radius $R^\infty \, > \, 16 \, Km \,
\frac{d}{120 \, pc}$~\cite{burwitz:2002}. In our previous
paper~\cite{cea:2003}, however, we suggested  that the seven data
points in the optical range could be interpreted  as  synchrotron
radiation~\cite{wallace:1977} emitted by electrons with energy
spectrum:
\begin{equation}
\label{syncroton}
N(E) \;  = \; \kappa \;  E^{-\eta} \; , \; \; \; 2. \, 10^{-3} \,
KeV \; \leq \; E
 \; \leq \; 7. \, 10^{-3} \; KeV  \; \; .
\end{equation}
Indeed, we find a quite good fit with $\eta \simeq 0.71$. If this
is the case the radiation in the optical range should display a
rather large linear polarization~\cite{wallace:1977}:
\begin{equation}
\label{polarization}
 \Pi \;  = \; \frac{\eta \; + \;1}{\eta \; +\; 7/3}
\; \simeq \; 0.56  \; .
\end{equation}
Interestingly enough, quite recently the distance measurement of
{\it {RXJ1856}} has been reassessed and it is now estimated to be
at $180 \; pc$ instead of $120 \; pc$~\cite{kaplan:2003}. In this
case, from Eq.~(\ref{rad-temp}) we obtain $ R^\infty = 6.6 \, Km$.
Assuming this value for $ R^\infty$ we can solve
Eq.~(\ref{radiuseff}). The result, displayed in Fig.~\ref{fig:01},
indicates that:
\begin{equation}
\label{rad-mass}
 R \, \simeq \, 5.0 \, Km \; \; , \; \; M \, \simeq \, 0.8 \, M_{\bigodot}
\; .
\end{equation}
It is worthwhile to stress that such a value for radius cannot be
accounted for within the neutron star theory. However, it must be
stressed that unusually small mass and radii are also obtainable
within the strange star theory (see, for instance
Ref.~\cite{Dey:1998,Li:1999,Sinha:2002}). \\
Even more, the new determination of the distance of {\it
{RXJ1856}} rules out the two blackbody interpretation of the
spectrum. Indeed, this model leads now to an effective radius
$R^\infty \, > \, 24 \, Km$, which is too large for a neutron
star. Thus, the new determination of the distance of {\it
{RXJ1856}} strongly supports our interpretation of the optical
emission. Moreover, we feel that alterative interpretations
recently proposed are problematic. Indeed, a different possibility
is to assume that the surface of the compact star is composed of a
solid matter~\cite{burwitz:2001,xu:2003,zane:2003,turolla:2003}.
Such a situation may occur at low temperatures  and high magnetic
fields~\cite{lai:1997,lai:2001}. Even though there are no reliable
calculations for this model, it has been suggested that it will
give an optical flux even lower than the blackbody
model~\cite{pavlov:2003}. We are lead, thus, to conclude that the
most realistic interpretation of the spectrum  of {\it {RXJ1856}}
is a thermal blackbody emission in $X$-ray and an optical electron
synchrotron radiation emission.
Let us consider the second nearest isolated compact star {\it
{RXJ0720}}, which has been detected by
ROSAT~\cite{haberl:1997,motch:1998} and observed with
XMM-Newton~\cite{paerels:2001,cropper:2001}. Remarkably, it turns
out that the spectrum of {\it {RXJ0720}} is almost identical to
that of {\it {RXJ1856}}. Indeed, {\it {RXJ0720}} exhibits a
blackbody $X$-ray spectrum with surface temperature $T^\infty \,
\simeq \, 80 \, eV$, a large X-ray to optical flux ratio, a low
$X$-ray luminosity, and an optical emission in excess over the
extrapolated $X$-ray blackbody. A recent
analysis~\cite{kaplan:2003a} of the optical, ultraviolet, and
$X$-ray data showed that the optical spectrum of {\it {RXJ0720}}
is not well fitted by a Rayleigh-Jeans tail, but it is best fitted
by a non thermal power law. We find, indeed, that the optical
spectrum can be interpreted  as synchrotron radiation emitted by
electrons with energy spectrum given by Eq.~(\ref{syncroton}) in
the relevant energy range. Moreover we obtain $\eta \simeq 0.63$
which is remarkably close to the value for {\it {RXJ1856}}. It
should be stressed, however, that in the fitting procedure we do
not consider the wavelength dependence of the interstellar
extinction~\cite{cardelli:1989,o'donnell:1994}. It turns out that
interstellar extinction leads to  somewhat larger values of
$\eta$, leading to a linear polarization which may reach about 90
$\%$. On the other hand, the authors of
Ref.~\cite{pavlov:2002,motch:2003} found that the best fit to the
$X$-ray spectrum on {\it {RXJ0720}} is provided by a blackbody
model with the effective radius and surface temperature:
\begin{equation}
\label{rad-temp-2}
 R^\infty \, \simeq \, (2.1 \,\pm \, 0.1) \; \frac{d}{100 \, pc} \, Km \; \; ,
 \; \; T^\infty \, \simeq
 \, 81 \, \pm \, 1 \, eV \; .
\end{equation}
The assumed distance to {\it {RXJ0720}} is $ d \simeq 300 \,
pc$~\cite{kaplan:2003a}. So that, from Eq.~(\ref{rad-temp-2}) we
infer $R^\infty \, \simeq \, 6.3 \, km$ , almost identical to the
{\it {RXJ1856}} effective radius. \\
We are led to conclude that the most realistic interpretation of
the emission spectra of {\it {RXJ1856}} and {\it {RXJ0720}}
indicates that these stars are rather compact with radius $R \,
\simeq \, 5.0 \, km$ and mass $M \, \simeq \, 0.8 \,
M_{\bigodot}$.
To complete our analysis we must check if the effective surface
temperature given by Eqs.~(\ref{rad-temp}) and (\ref{rad-temp-2})
are compatible with the P-star cooling curves. As discussed in
Ref.~\cite{cea:2003}, we assume stars of uniform density and
isothermal. The equation which determines the thermal history of a
P-star is:
\begin{equation}
\label{cooling}
 C_V \; \frac{d T}{d t}
\; = \; - \; (L_{\nu} \; + \; L_{\gamma}) \; ,
\end{equation}
where $L_{\nu}$ is the neutrino luminosity, $L_{\gamma}$ is the
photon luminosity and $C_V$ is the specific heat. Assuming
blackbody photon emission from the surface at an effective surface
temperature $T_S$ we have:
\begin{equation}
\label{blackbody}
 L_{\gamma} \; = \; 4 \, \pi \, R^2 \, \sigma_{SB} \, T_S^4 \; ,
\end{equation}
where $\sigma_{SB}$ is the $Stefan-Boltzmann$ constant. In
Ref.~\cite{cea:2003} we assumed that the surface and interior
temperature were related by:
\begin{equation}
\label{surface}
 \frac{T_S}{T} \; = \; 10^{-2} \; a \; \; , \; \; 0.1 \; \lesssim a \; \lesssim 1.0
 \; .
\end{equation}
\begin{figure}
\includegraphics[width=0.90\textwidth,clip]{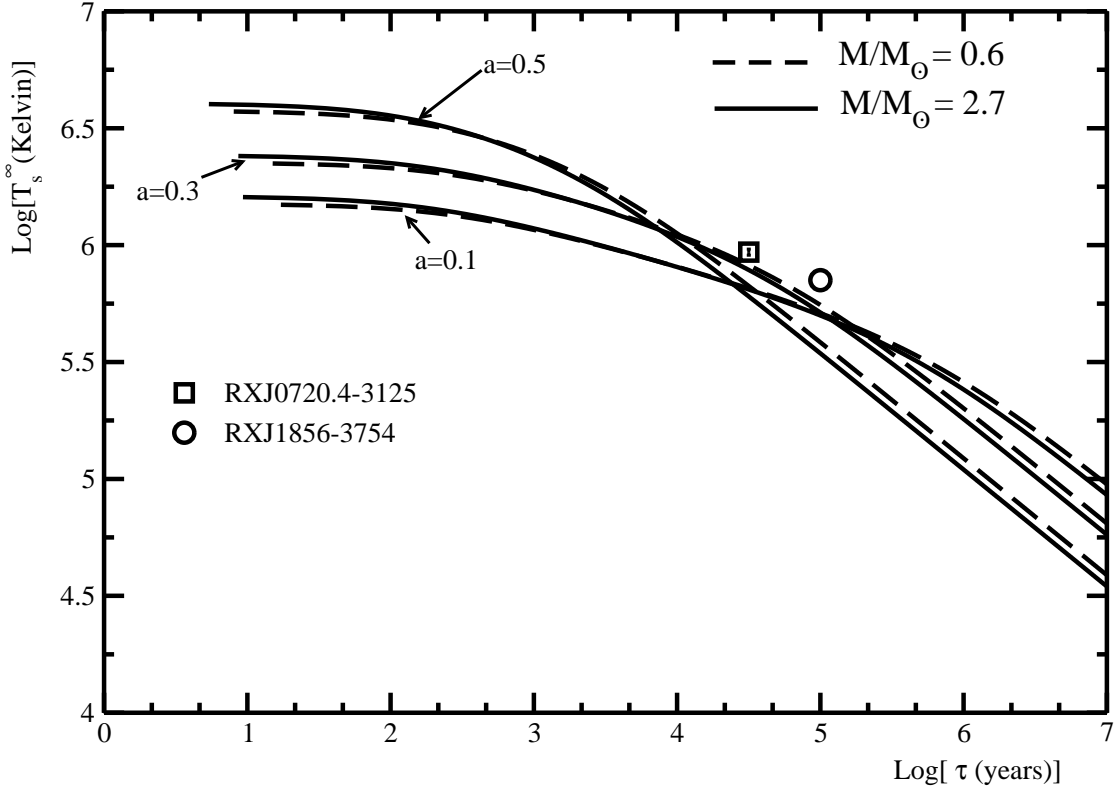}
\caption{\label{fig:02}
P-star cooling curves for $M = 0.6 \, M_{\odot}$ and $M = 2.7 \,
M_{\odot}$  and three different values of the parameter $a$ in
Eq.~(\ref{surface}). }
\end{figure}
Equation~(\ref{surface}) is relevant for a P-star which is not
bare, namely for P-stars which are endowed with a thin crust. It
turns out that, like strange
stars~\cite{alcock:1986a,kettner:1995} (see also the discussion in
Ref.~\cite{glendenning:2000}, pag.~428), P-stars have a sharp edge
of thickness of the order of about $ 1 \; fermi$. On the other
hand, electrons which are bound by the coulomb attraction, extend
several hundred $fermis$ beyond the edge. As a consequence, on the
surface of the star there is a positively charged layer which is
able to support a thin crust of ordinary matter (most probably,
atomic hydrogen). Thus we see that the vacuum gap between the core
and the crust, which is of order of hundred $fermis$, leads to a
strong suppression of the surface temperature with respect to the
core temperature. The precise relation between $T_S$ and $T$ could
be obtained by a careful study of the crust and core thermal
interaction. In the case of neutron stars this study has been
performed in Ref.~\cite{gudmundsson:1983}. In any case, our
phenomenological relation Eq.~(\ref{surface}) allows a wide
variation of $T_S$, which encompasses the neutron star relation of
Ref.~\cite{gudmundsson:1983}. Moreover, our cooling curves display
a rather weak dependence on the parameter $a$ in
Eq.~(\ref{surface}) only for stellar age up to $ \tau \, \sim \,
10^3 \, years $ (see Fig.~\ref{fig:02}). \\
The neutrino luminosity $L_{\nu}$ in Eq.~(\ref{cooling}) is given
by the direct $\beta$-decay quark
reactions~\cite{iwamoto:1980,burrows:1980}, the dominant cooling
processes by neutrino emission. It turns out that the neutrino
luminosity is~\cite{cea:2003}:
\begin{equation}
\label{luminosity}
 L_{\nu} \; \simeq \; 3.18 \; 10^{36} \; \frac{erg}{s} \; T_9^8 \;
 \frac{M}{M_{\bigodot}} \; \frac{\varepsilon_0}{\varepsilon} \;
   \frac{\sqrt{gH}}{1 \, GeV} \;
 \;  ,
\end{equation}
where $T_9$ is the temperature in units of $10^9$ $\, {}^\circ K$,
and $\varepsilon_0 = 2.51 \; 10^{14} gr/cm^3$ is the nuclear
density. Note that the neutrino luminosity $L_{\nu}$ has the same
temperature dependence as the neutrino luminosity by  modified
URCA reactions in neutron stars (see, for instance
Ref.~\cite{shapiro:1983}), but it is more than two order of
magnitude smaller.
The specific heat is given by:
\begin{equation}
\label{specific}
 C_V \; \simeq \; 0.92 \; 10^{55} \;  T_9 \;
 \frac{M}{M_{\bigodot}} \; \frac{\varepsilon_0}{\varepsilon} \;
   (\frac{\sqrt{gH}}{1 \, GeV})^2 \; ,
\end{equation}
which in physical units reads:
\begin{equation}
\label{specific-phys}
 C_V \; \simeq \; 1.27 \; 10^{39} \; \frac{erg}{{}^\circ K} \: T_9 \;
 \frac{M}{M_{\bigodot}} \; \frac{\varepsilon_0}{\varepsilon} \;
   (\frac{\sqrt{gH}}{1 \, GeV})^2 \; .
\end{equation}
From Eq.~(\ref{specific-phys}) we see that the P-star specific
heat is of the same order of the neutron star specific
heat~\cite{shapiro:1983}.
In Figure~\ref{fig:02} we report our cooling curves, obtained by
integrating Eq.~(\ref{cooling}) with $T^{(i)}_9 = 1.4$. It is
worthwhile to note that the effective surface temperature almost
does not depends on the star mass. Indeed, even though the
luminosity and the energy density vary by about an order of
magnitude when  the stellar mass $M/M_{\odot}$ ranges from $0.6$
to $2.7$, the derivative of the temperature depends only on the
ratio of the luminosity to the specific heat. It turns out that
this ratio varies by less than a factor two in the above range of
masses. \\
In our previous paper~\cite{cea:2003} we compared the cooling
curves with available pulsar data and found a quite satisfying
agreement. In Figure~\ref{fig:02} we report the effective surface
temperatures of {\it {RXJ1856}} and {\it {RXJ0720}}. The ages of
the two compact stars have been fixed by matching with the cooling
curves. In this way we obtain:

\begin{equation}
\label{star-age}
\begin{split}
 \tau & \; \simeq \; \; \; 10^5 \; \; \; years \; \; \; \; \; \; \; \;
 for \; \; \; \; RXJ1856 \; , \\
 \tau & \; \simeq \; 3.2 \; 10^4 \;  years  \; \; \; \; \; \;
 for \; \; \; \; RXJ0720  \; .
\end{split}
\end{equation}
In absence of a measured period and period derivative our age
estimate for {\it {RXJ1856}} is a challenge to future
observations. On the other hand, recent timing analysis  performed
in Ref.~\cite{zane:2002,zane:2002a} allowed to determine the
period and period derivative of {\it {RXJ0720}} :
\begin{equation}
\label{p-pdot}
 P \; \simeq \; 8.391 \; \; s \; \; \; \;  , \; \; \; \;
 \dot{P}  \; \simeq \; 5.41 \;  10^{-14}
 \; \; \; \frac{s}{s} \; \; \; ,
\end{equation}
where for definiteness we use solution (1) of Table~2 in
Ref.~\cite{zane:2002a}. From the period and period derivative we
may estimate the age of the star. Indeed, assuming  star slowdown
by dipolar magnetic braking, one obtains the age and the surface
magnetic field:
\begin{equation}
\label{spin-age}
 \tau\; = \; \frac{P}{2 \, \dot{P}} \; \; \left [ 1 \, - \,
 \left ( \frac{P_0}{P}\right )^2 \right ]  \; \; \; ,
\end{equation}
\begin{equation}
\label{mag-surf}
 B_S \; \simeq \; \sqrt{\frac{3 \, I \, P \, \dot{P}}{8 \, \pi^2 \, R^6}}  \; \; \; ,
\end{equation}
where $P_0$ is the initial period, $I$ is the moment of inertia.
Assuming that the pulsar started out life rotating much faster
than present, then $P_0 \, \ll \, P$ and Eq.~(\ref{spin-age})
reduces to the characteristic age $\tau_c$, Eq.~(\ref{char-age}).
Using Eq.~(\ref{p-pdot}) we get the   characteristic age for {\it
{RXJ0720}}:
\begin{equation}
\label{char-age-0720}
 \tau_c \; \simeq \; 2.5 \; 10^{6} \; years \; \;,
\end{equation}
which is about a factor $10^2$ greater than our estimate in
Eq.~(\ref{star-age}). Note, however, that the characteristic age
gives only an upper limit to the true age $\tau$,
Eq.~(\ref{spin-age}), which can be significantly smaller than
$\tau_c$ if the initial period $P_0$ is close to $P$. As a matter
of fact, within the neutron star theory it is believed that
pulsars are generally born as rapid rotators, so that they
generate a sizeable magnetic field by the dinamo mechanism.  On
the other hand, the magnetic field of {\it {RXJ0720}} is quite
large. Indeed, using Eq.~(\ref{mag-surf}) together with the mass,
radius, period and period derivative values given in
Eq.~(\ref{rad-mass}) and Eq.~(\ref{p-pdot}) respectively, we get:
\begin{equation}
\label{mag-surf-comp}
 B_S \; \simeq \; 9.4 \; 10^{19} \;  \sqrt{ P \, \dot{P}} \; \; Gauss  \; \simeq
 \;  6.3 \; 10^{13} \; \; Gauss \; .
\end{equation}
In Ref.~\cite{cea:2004}, where we  discuss the generation of
magnetic field in P-stars, we argue that huge magnetic field
requires that P-stars must born rotating slowly. So that we see
that, in the case of {\it {RXJ0720}}, our estimate of the magnetic
field Eq.~(\ref{mag-surf-comp}) implies a rather large value of
$P_0$, in fact comparable to $P$. It is easy to see that, if
\begin{equation}
\label{P_0-0720}
 P_0 \; \simeq \; 8.336 \; \;sec \; \;,
\end{equation}
then Eq.~(\ref{spin-age}) gives $\tau \, \simeq \, 3.2 \, 10^4 \,
years$, in perfect agreement with Eq.~(\ref{star-age}). Moreover,
we shall show in Ref.~\cite{cea:2004} that in this way we may
solve completely the puzzling discrepancy between the
characteristic age and the true age inferred from the associated
supernova remnants in the case of anomalous $X$-ray pulsars {\it
1E 2259+586} and {\it 1E 1841-045}.
Interestingly enough, our estimate of the magnetic field agrees
with a recent analysis of the spectrum of {\it
{RXJ0720}}~\cite{haberl:2003a}. The authors of
Ref.~\cite{haberl:2003a} performed a spectral analysis of four
XMM-Newton observations of  {\it {RXJ0720}}. They find deviations
in the spectra from a planckian shape which are interpreted as an
absorption line. From the pulse-phase averaged spectra they derive
for a gaussian-shaped line :
\begin{equation}
\label{line-0-0720}
 E \; \simeq \; 271 \; \;eV \; \;, \; \; \sigma \; \simeq \; 64 \; \;eV
 \;,
\end{equation}
and an equivalent width of $-40 \; eV$. It is natural to assume
that cyclotron resonance absorption are likely the origin for the
absorption feature seen in the spectra. If we restrict to
electrons and protons as the origin of this absorption, we get:
\begin{equation}
\label{cyclotr-line}
 E_{B_e} \; \simeq \; 11.6 \; \; B_{12}  \; \; KeV  \; \; \; \; , \; \; \; \;
 E_{B_p} \; \simeq \; 6.3 \; \;  B_{12}  \; \; eV \; \; .
\end{equation}
From the magnetic field Eq.~(\ref{mag-surf-comp}), it is clear
that electrons are excluded as the origin of the cyclotron line.
On the other hand, in the case of protons, using
\begin{equation}
\label{zeta}
 \sqrt{ 1 \; - \; \frac{2 \, G \, M}{R}}  \; \simeq \;
0.726 \; \; \; ,
\end{equation}
we obtain the proton cyclotron line at:
\begin{equation}
\label{proton-line}
 E_{p} \; \simeq \; 288 \; \;  eV
\end{equation}
in striking agreement with Eq.~(\ref{line-0-0720}).

The last point which we would like to comment is the general
feature of the emission spectrum which seems to be consistent with
a soft $X$-ray  thermal blackbody emission from the surface and
synchrotron (or, more generally, power law) optical and UV
radiation. As it is well known, in general pulsar emission is
powered by the rotational energy:
\begin{equation}
\label{ener-rot}
  E_{R} \; = \;  \frac{1}{2} \; I \; \; \omega^2 \; \; \; .
\end{equation}
Thus, the spin-down power output is given by:
\begin{equation}
\label{ener-rot-dot}
 - \; \dot{E}_{R} \; = \; - \; I \; \; \omega \; \dot{\omega} \; \; = \; 4 \;
 \pi^2 \; I \; \frac{\dot{P}}{P^3} \; .
\end{equation}
On the other hand, an important source of energy is provided by
the magnetic field. Indeed, the classical energy stored into the
magnetic field is:
\begin{equation}
\label{ener-mag}
 E_{B} \; = \; \int_{r \, \geq \, R} \; \frac{1}{8 \, \pi} \; B^2(r) \; \;  \; \; ,
\end{equation}
Assuming a dipolar magnetic field:
\begin{equation}
\label{dip-mag}
 B(r) \; = \;  B_S \; \; \left ( \frac{R}{r} \right )^3 \; \; for \; \; \;
 r \, \geq \, R \; \;  \; \; ,
\end{equation}
Eq.~(\ref{ener-mag}) leads to:
\begin{equation}
\label{ener-mag-dip}
 E_{B} \; = \; \frac{1}{6 } \; B_S^2 \; \; R^3 \; \; .
\end{equation}
As discussed in Section~1, the surface magnetic field turns out to
be time dependent. So that, using Eq.~(\ref{mag-time}) the
magnetic power output is given by:
\begin{equation}
\label{ener-mag-dot}
 \dot{E}_{B} \; = \; \frac{2}{3} \;  B_0^2 \; \; R^3 \; \;
  \; \frac{\dot{P}}{P} \; .
\end{equation}
For rotation-powered pulsars it turns out that $ \dot{E}_{B} \;
\ll \; - \, \dot{E}_{R}$. However, if the dipolar magnetic field
is strong enough, then the magnetic power can be of the order, or
even greater than the spin-down power. In Ref.~\cite{cea:2004} we
will show that this fact leads to  the pulsar death line, which is
the line that in the $P-\dot{P}$ plane separates rotation-powered
pulsars from magnetic-powered pulsars. \\
In the case of {\it {RXJ0720}}, by using Eqs.~(\ref{rad-mass})
and~(\ref{mag-surf-comp}), we get:
\begin{equation}
\label{ener-rot-dot-comp}
- \; \dot{E}_{R} \; \simeq \; 0.9 \; 10^{44} \; \; erg \; \;  \;
\frac{\dot{P}}{P} \; ,
\end{equation}
and
\begin{equation}
\label{ener-mag-dot-comp}
\dot{E}_{B} \; \simeq \; 0.3 \; 10^{44} \; \; erg  \; \; \;
\frac{\dot{P}}{P} \; \; .
\end{equation}
We see that, within our uncertainty, $ \dot{E}_{B} + \;
\dot{E}_{R} \; \simeq \; 0$, namely almost all the rotation energy
is stored into the increasing magnetic field. As a consequence,
the emission from the star consists in thermal blackbody radiation
form the surface. In addition, it could eventually also be present
 a faint synchrotron emission superimposed to the thermal
radiation. We believe that {\it {RXJ1856}} is exactly in this
state. On the other hand, the energy stored into the magnetic
field can be released if the star undergoes a glitch. Indeed, as
thoroughly discussed in Ref.~\cite{cea:2004}, glitches originate
from dissipative effects in the inner  core of the star leading to
a decrease of the strength of the dipolar magnetic field. So that,
soon after the glitch there is a release of magnetic energy. It is
remarkable that this picture is consistent with the recently
detected long-term variability in the $X$-ray emission of {\it
{RXJ0720}}~\cite{deVries:2004}. As a consequence we predict that a
similar emission  could also be detected  from {\it {RXJ1856}}
when and if this compact star will suffer a glitch.
\renewcommand{\thesection}{\normalsize{{\arabic{section}}}}
\section{\normalsize{CONCLUSIONS}}
\renewcommand{\thesection}{\arabic{section}}
In this paper we have discussed in details the nearest isolated
radio quiet pulsar {\it {RXJ1856}} and {\it {RXJ0720}}. Our
results show that these stars are most likely P-stars with   $M \,
\simeq 0.8 \, M_{\bigodot}$ and $R \, \simeq 5 \, Km$. In a
forthcoming paper we will discuss the generation of magnetic field
and the glitch mechanism in P-stars. Moreover, we will show that
our results compare rather well with available pulsar data.
In conclusion, we feel  that the ability of our P-star theory for
accounting in a coherent fashion several observational features of
pulsars should suggest that it warrants serious consideration.


\begin{thebibliography}{99}
%
\bibitem{hewish:1968}
A.~Hewish, S.~G.~Bell, J.~D.~H.~Pilkington, P.~F.~Scott, and
R.~A.~Collins, Nature\ {\bf 217}, 709 (1968).
%

\bibitem{pacini:1968}
F.~Pacini,  Nature\ {\bf 219}, 145 (1968).
%
\bibitem{gold:1968}
T.~Gold,  Nature\ {\bf 218}, 731 (1968).
%
%
\bibitem{michel:1982}
See, for instance, F.~C.~Michel,  Rev. \ Mod. \ Phys. \ {\bf 54},
1 (1982); F.~C.~Michel, {\em The State of Pulsar Theory },
astro-ph/0308347.
%
%
\bibitem{michel:1991}
F.~C.~Michel,
\newblock {\em Theory of Neutron Star Magnetospheres},
(The University of Chicago Press, Chicago, 1991).
%
%
\bibitem{goldreich:1969}
P.~Goldreich and W.~H.~Julian,  Astrophys.\ J.\  {\bf 157}, 869
(1969).
%
%
\bibitem{sturrock:1971}
P.~A.~Sturrock,  Astrophys.\ J.\  {\bf 164}, 529 (1971).
%
%
\bibitem{cea:2003}
P.~Cea, {\em P-Stars}, astro-ph/0301578.
%
%
\bibitem{bethe:1990}
For a review, see: H.~A.~Bethe, Rev.\ Mod.\ Phys.\  {\bf 62}, 801
(1990).
%
%
\bibitem{cea:2004}
P.~Cea, {\em Magnetic Fields and Glitches in P-Stars}, in
preparation.
%
%
\bibitem{glendenning:2000}
See, for instance: N~K.~Glendenning,
\newblock {\em Compact Stars: Nuclear Physics, Particle Physics,
and General Relativity}, 2nd ed. (Springer-Verlag New York, 2000).
%
%
\bibitem{alcock:1986}
C.~Alcock, E.~Farhi and A.~Olinto, Astrophys.\ J.\  {\bf 310}, 261
(1986).
%
\bibitem{haensel:1986}
P.~Haensel, J.~L.~Zdunik and R.~Schaeffer, Astron.\ Astrophys.\
{\bf 160}, 121 (1986).
%
%
%
\bibitem{bodmer:1971}
A.~R.~Bodmer, Phys. Rev. {\bf D 4}, 1601 (1971).
%
\bibitem{witten:1984}
E.~Witten, Phys. Rev. {\bf D 30}, 272 (1984).
%
%
\bibitem{farhi:1984}
E.~Farhi and R.~L.~Jaffe, Phys.\ Rev.\ D {\bf 30}, 2379 (1984).
%
%
\bibitem{walter:1996}
F.~M.~Walter, S.~J.~Wolk, and R.~Neuh\"{a}user, Nature {\bf 379},
233 (1996).
%
%
%
\bibitem{neuhauser:1997}
R.~Neuh\"{a}user, H.-C.~Thomas, R.~Danner, S.~Peschke, and
F.~M.~Walter, Astron.\ Astrophys.\ {\bf 318}, L43 (1997).
%
%
\bibitem{haberl:2003}
F.~Haberl, {\em  AXPs and X-ray dim neutron stars: Recent
XMM-Newton and Chandra results}, arXiv:astro-ph/0302540.
%
%
%
\bibitem{pavlov:2003}
G.~G.~Pavlov and V.~E.~Zavlin, {\em Thermal Radiation from Cooling
Neutron Stars}, arXiv:astro-ph/0305435.
%
%
\bibitem{burwitz:2002}
V.~Burwitz, F.~Haberl, R.~Neuh\"{a}user, P.~Predehl,
J.~Tr\"{u}mper, and V.~E.~Zavlin, {\em The thermal radiation of
the isolated neutron star RX J1856.5-3754 observed with Chandra
and XMM-Newton}, astro-ph/0211536 .
%
%
\bibitem{wallace:1977}
See, for instance:
%
W.~H.~Wallace,
\newblock {\em Radiation Processes in Astrophysics}
(MIT Press, Cambridge, 1977);
%
V.~L.~Ginzburg,
\newblock {\em Theoretical Physics and Astrophysics}
(Pergamon, Oxford, 1979).
%
%
%
\bibitem{kaplan:2003}
D.~Kaplan,
\newblock {\em Optical Observations of Isolated Neutron Stars},
contribute to the workshop {\em Physics and Astrophysics of
Neutron Stars}, July 28 - August 1, 2003, Santa Fe, New Mexico.
%
%
\bibitem{Dey:1998}
M.~Dey, I.~Bombaci, J.~Dey, S.~Ray and B.~C.~Samanta, Phys.\
Lett.\ B {\bf 438}, 123 (1998)  [Addendum-ibid.\ B {\bf 447}, 352
(1999) ].
%
%
%
\bibitem{Li:1999}
X.~D.~Li, I.~Bombaci, M.~Dey, J.~Dey and E.~P.~J.~van den Heuvel,
Phys.\ Rev.\ Lett.\  {\bf 83}, 3776 (1999) .
%
%
%
\bibitem{Sinha:2002}
M.~Sinha, J.~Dey, M.~Dey, S.~Ray and S.~Bhowmick, Mod.\ Phys.\
Lett.\ A {\bf 17}, 1783 (2002).
%
%
\bibitem{burwitz:2001}
V.~Burwitz, V.~E.~Zavlin, R.~Neuhaeuser, P.~Predehl, J.~
Tr\"umper, and A.~C.~Brinkman, Astron.\ Astrophys.\ {\bf 379}, L35
(2001).
%
%
%
\bibitem{xu:2003}
R.~X.~Xu, {\em Solid bare strange quark stars}, astro-ph/0310050.
%
\bibitem{zane:2003}
S.~Zane, R.~Turolla and J.~J.~Drake, {\em RX J1856.5-3754: Bare
quark star or naked neutron star?}, astro-ph/0302197.
%
\bibitem{turolla:2003}
R.~Turolla, S.~Zane and J.~J.~Drake, {\em Bare quark stars or
naked neutron stars? The case of RX J1856.5-3754},
astro-ph/0308326.
%
\bibitem{lai:1997}
D.~Lai and  E.~E.~Salpeter, Astrophys.\ J.\  {\bf 491}, 270
(1997).
%
\bibitem{lai:2001}
D.~Lai, Rev.\ Mod.\ Phys.\ {\bf 73}, 629 (2001).
%
%
%
\bibitem{haberl:1997}
F.~Haberl, C.~Motch,  D.~A.~H.~Buckley, F.-J.~Zickgraf, and
W.~Pietsch, Astron.\ Astrophys.\ {\bf 326}, 662 (1997).
%
\bibitem{motch:1998}
C.~Motch and F.~Haberl, Astron.\ Astrophys.\ {\bf 333}, L59
(1998).
%
%
\bibitem{paerels:2001}
F.~Paerels, C.~Motch, F.~Haberl, V.~E.~Zavlin, S.~Zane, G.~Ramsay,
 M.~Cropper, and B.~Brinkman, Astron.\ Astrophys.\ {\bf 365}, 302
 (2001).
%
\bibitem{cropper:2001}
M.~Cropper, S.~Zane, G.~Ramsay, F.~Haberl and C.~Motch, Astron.\
Astrophys.\ {\bf 365}, L302 (2001).
%
\bibitem{kaplan:2003a}
D.~L.~Kaplan, M.~H.~van Kerkwijk, H.~L.~Marshall,  B.~A.~Jacoby,
S.~R.~Kulkarni, and  D.~A.~Frail,  Astrophys.\ J.\  {\bf 590},
1008 (2003).
%
%
\bibitem{cardelli:1989}
J.~A.~Cardelli, G.~C.~Clayton and  J.~S.~Mathis,  Astrophys.\ J.\
{\bf 345}, 245 (1989).
%
%
\bibitem{o'donnell:1994}
J.~E.~O'Donnell,  Astrophys.\ J.\  {\bf 422}, 158 (1994).
%
\bibitem{pavlov:2002}
G.~G.~Pavlov, V.~E.~Zavlin and D.~Sanwal, in  {\em Neutron Stars,
Pulsar and Supernova}, ed. W.~Beckerm H.~Lesch and J.~Tr\"umper,
(MPE-Report-278, 2002), p.~273.
%
\bibitem{motch:2003}
C.~Motch, V.~E.~Zavlin and F.~Haberl, {\em The proper motion and
energy distribution of the isolated neutron star RX J0720.4-3125},
astro-ph/0305016.
%
%
\bibitem{alcock:1986a}
C.~Alcock, E.~Farhi, and A.~V.~Olinto,  Astrophys.\ J.\  {\bf
310}, 261 (1986).
%
%
\bibitem{kettner:1995}
Ch.~Kettner, F.~Weber, M.~K.~Weigel, and N.~K.~Glendenning, Phys.\
Rev.\ D\ {\bf 51}, 1440 (1995).
%
%
\bibitem{gudmundsson:1983}
E.~H.~Gundmundsson, C.~J.~Pethick, and R.~I.~Epstein,  Astrophys.\
J.\ {\bf 272}, 286 (1983).
%
%
\bibitem{iwamoto:1980}
N.~Iwamoto, Phys.\ Rev.\ Lett.\ {\bf 44}, 1637 (1980).
%
\bibitem{burrows:1980}
A.~Burrows, Phys.\ Rev.\ Lett.\ {\bf 44}, 1640 (1980).
%
%
\bibitem{shapiro:1983}
S.~L.~Shapiro and S.~A.~Teukolsky,
\newblock {\em Black Holes, White Dwarfs, and Neutron Stars},
(John Wiley \& Sons , 1983).
%
%
\bibitem{zane:2002}
S.~Zane, F.~Haberl, M.~Cropper, V.~E.~Zavlin, D.~Lumb, S.~Sembay
and C.~Motch, Mon.\ Not.\ Roy.\ Astron.\ Soc.\  {\bf 334}, 345
(2002).
%
%
\bibitem{zane:2002a}
S.~Zane, F.~Haberl M.~Cropper, V.~Zavlin, D.~Lumb, S.~Sembay, and
C.~Motch, {\em Timing Analysis of the Isolated Neutron Star RX
J0720.4-3125}, astro-ph/0203111.
%
%
\bibitem{haberl:2003a}
F.~Haberl, V.~E.~Zavlin, J.~Truemper, and V.~Burwitz,
\newblock
{\em A phase-dependent absorption line in the spectrum of the
X-ray pulsar RX J0720.4-3125}, astro-ph/0312413.
%
%
\bibitem{deVries:2004}
C.~P.~de Vries, J.~Vink, M.~Mendez, and F.~Verbunt,
\newblock
{\em Long-term variability in the X-ray emission of RX
J0720.4-3125}, astro-ph/0401028.
%
%
\end{thebibliography}
\end{document}